\documentclass[prb,twocolumn,showpacs,superscriptaddress]{revtex4-2}
\usepackage{epsfig}
\usepackage{amsmath,amssymb}
\usepackage{graphicx}
\usepackage{color}
\usepackage{braket}
\usepackage{dcolumn}

\usepackage[colorlinks,breaklinks,bookmarks=true,citecolor=blue,linkcolor=red,urlcolor=blue]{hyperref}

\newcommand{\nabaco}{Na$_2$BaCo(PO$_4$)$_2$}
\newcommand{\jhalf}{\ensuremath{\tilde{j}_{\text{eff}}=\frac12}}
\newcommand{\ltilde}{\ensuremath{\widetilde{\lambda}}}
\newcommand{\im}{\text{i}}
\newcommand{\ts}[1]{_\text{#1}}

\begin{document}

\title{Frustration enhanced by Kitaev exchange in a  $\boldsymbol{\tilde{j}_{\text{eff}}=\frac12}$ triangular antiferromagnet}

\author{C.~Wellm}
\affiliation{Leibniz Institute for Solid State and Materials Research IFW Dresden, 01171 Dresden, Germany}
\affiliation{Institute for Solid State and Materials Physics, TU Dresden, 01069 Dresden, Germany}

\author{W.~Roscher}
\affiliation{Leibniz Institute for Solid State and Materials Research IFW Dresden, 01171 Dresden, Germany}

\author{J.~Zeisner}
\affiliation{Leibniz Institute for Solid State and Materials Research IFW Dresden, 01171 Dresden, Germany}
\affiliation{Institute for Solid State and Materials Physics, TU Dresden, 01069 Dresden, Germany}

\author{A.~Alfonsov}
\affiliation{Leibniz Institute for Solid State and Materials Research IFW Dresden, 01171 Dresden, Germany}

\author{R.~Zhong}
\affiliation{Department Of Chemistry, Princeton University, Princeton, NJ 08544, USA}

\author{R.~J.~Cava}
\affiliation{Department Of Chemistry, Princeton University, Princeton, NJ 08544, USA}

\author{A.~Savoyant}
\affiliation{Aix-Marseille  Univ.,  CNRS,  IM2NP-UMR  7334,  13397  Marseille  Cedex  20,  France}

\author{R.~Hayn}
\affiliation{Leibniz Institute for Solid State and Materials Research IFW Dresden, 01171 Dresden, Germany}
\affiliation{Aix-Marseille  Univ.,  CNRS,  IM2NP-UMR  7334,  13397  Marseille  Cedex  20,  France}

\author{J.~van~den~Brink}
\affiliation{Leibniz Institute for Solid State and Materials Research IFW Dresden, 01171 Dresden, Germany}
\affiliation{W\"urzburg-Dresden Cluster of Excellence ct.qmat, TU Dresden, D-01062 Dresden, Germany}

\author{B.~B\"uchner}
\affiliation{Leibniz Institute for Solid State and Materials Research IFW Dresden, 01171 Dresden, Germany}
\affiliation{W\"urzburg-Dresden Cluster of Excellence ct.qmat, TU Dresden, D-01062 Dresden, Germany}

\author{O.~Janson}
\affiliation{Leibniz Institute for Solid State and Materials Research IFW Dresden, 01171 Dresden, Germany}

\author{V.~Kataev}
\affiliation{Leibniz Institute for Solid State and Materials Research IFW Dresden, 01171 Dresden, Germany}

\date{\today}

\begin{abstract}
Triangular Heisenberg antiferromagnets are prototypes of geometric frustration,
even if for nearest-neighbor interactions quantum fluctuations are not usually
strong enough to destroy magnetic ordering: stronger frustration is required to
stabilize a spin-liquid phase. On the basis of static magnetization and electron spin resonance
measurements, we demonstrate the emergence of $\jhalf$ moments in the
triangular-lattice magnet Na$_2$BaCo(PO$_4$)$_2$. These moments are subject to
an extra source of frustration that causes magnetic correlations to set in far
above both the magnetic ordering and Weiss temperatures. Corroborating the
$\jhalf$ ground state, theory identifies ferromagnetic Kitaev exchange
anisotropy as an additional frustrating agent, altogether putting forward
Na$_2$BaCo(PO$_4$)$_2$ as a promising Kitaev spin-liquid material.
\end{abstract}

\maketitle

\section{Introduction}
Quantum spin liquids are of topical interest as a promising host for
fractionalized quasiparticles~\cite{balents10, savary16, zhou17, knolle19,
broholm20}. The quest for a spin-liquid ground state goes back to Anderson's
conjecture on resonating valence bonds in the geometrically frustrated
two-dimensional triangular lattice~\cite{anderson73}. It is now firmly
established, however, that the pure spin 1/2 triangular-lattice Heisenberg
model has an ordered ground state~\cite{bernu92, *bernu93, *bernu94, *white07}.
This seemingly frustrating result is a blessing in disguise, though, as
subsequent studies of modified triangular-lattice models found that long-range
ordering can be suppressed by three disparate alterations: additional
second-neighbor exchange~\cite{iqbal16, bauer17, saadatmand17, gong17,
ferrari19, hu19}, spatial anisotropy~\cite{yunoki06, ghorbani16}, and exchange
anisotropy~\cite{zhu18, maksimov19}.  These findings opened new perspectives
for realizing a spin-liquid state in the triangular lattice geometry, and
boosted the search for candidate materials.

\begin{figure}[tb]
    \includegraphics[width=\columnwidth]{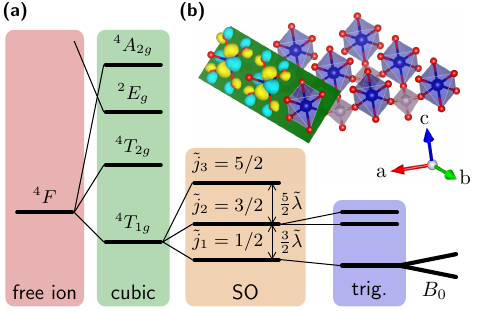}
    \caption{(a) Energy diagram showing the splitting of the $^4F$ state of
Co$^{2+}$ in a cubic crystal field, the SO coupling, trigonal
distortion and the Zeeman splitting due to the magnetic field $B_0$ in the
$^4T_{1g}$ multiplet. (b) Fragment of a triangular layer in the \nabaco\
crystal structure, with CoO$_6$ octahedra bridged by PO$_4$ tetrahedra. For a
pair of neighboring Co atoms, the leading contribution to hopping stems from
the depicted Wannier functions. Note that the respective Co $d$ orbitals lie in
the same plane.}
    \label{fig:overview}
\end{figure}

A recent breakthrough in triangular-lattice magnets pertains to the discovery
of two material classes~\cite{li20jpcm}. The first group of candidate materials
are $4f$ magnets with Yb$^{3+}$: YbMgGaO$_4$~\cite{li15scirep, li15prl} and
NaYb$X_2$ (with $X$=O,S,Se)~\cite{liu18}. The key element of their physics is
the crystal-field (CF) splitting of Yb$^{3+}$, giving rise to the ground-state
doublet and hence the effective spin 1/2 behavior at low temperatures, and, in
the former material, the significant effects of {chemical disorder between the
magnetic layers}~\cite{Paddison2017}. A similar scenario, with the lowest-lying
doublet dominating the low-temperature physics, is realized in the other
promising group of materials -- Co$^{2+}$ cobaltates. Here, the ground-state
doublet is stabilized by the spin-orbit (SO) coupling as long as the low-symmetry
crystal field distortions remain small (Fig.~\ref{fig:overview}, a).  Despite
the overall tendency towards ferromagnetic exchange, several cobaltates with
antiferromagnetic (AFM) interactions are known. The most studied material of this
handful, Ba$_3$CoSb$_2$O$_9$, shows intriguing high-field
behavior~\cite{shirata12, susuki13, koutroulakis15}, but evades the spin-liquid
regime by developing a long-range order at 3.7\,K~\cite{shirata12}.
Interestingly, in the related layered honeycomb magnet BaCo$_2$(AsO$_4$)$_2$
the magnetic order can be suppressed by a low magnetic field yielding a
nonmagnetic material~\cite{Zhong2020}.

The recently synthesized cobaltate \nabaco\ (Fig.~\ref{fig:overview}, b) is a
promising candidate for realization of a spin-liquid state on a triangular
lattice. The first experimental study reported a sizable antiferromagnetic
exchange and no long-range ordering down to 50\,mK~\cite{zhong19}.  Very
recently, ultralow-temperature specific heat measurements revealed a magnetic
ordering transition at $T_{\rm N} = 148$\,mK~\cite{li20natcommun},
i.e.,$\sim20$ times smaller than the Weiss temperature $\theta = 2.5$\,K
obtained from the static magnetic susceptibility $\chi (T)$
\cite{li20natcommun}. Another interesting observation is the linear dependence
of the residual thermal conductivity~\cite{Lee21}, compatible with the
spinon Fermi surface scenario advocated for critical spin liquids. However, the
microscopic model for \nabaco\ remains hitherto unexplored.

In this Letter, we fill this gap and demonstrate that exchange anisotropy
stabilizes the spin-liquid behavior in \nabaco\ over a temperature range
spanning two orders of magnitude. Our  Co$^{2+}$ electron spin resonance (ESR)
measurements firmly establish the effective spin 1/2 state of the Co$^{2+}$
ions~\cite{Note1} by determining the $g$~factor tensor and evaluating the
excitation energy of the first excited state with effective spin 3/2. Further,
the ESR data indicate the onset of magnetic correlations at the surprisingly
high temperature of $\sim20$\,K, which is two orders of magnitude larger than
$T_{\rm N}$.  Hence, magnetic correlations in this material build up far above
the phenomenological energy scale set by the moderate Weiss temperature and the
low saturation field of the magnetization. To reconcile these seemingly
conflicting observations, we perform a microscopic analysis based on
density-functional-theory (DFT) band-structure and multiplet calculations. We
confirm the effective spin 1/2 (hereafter \jhalf) behavior, estimate
theoretically the $g$~factors, and justify the restriction of the magnetic
system to the strong dominance of the nearest-neighbor exchange $J_1$, which
however turns out to be substantially larger than previous experimental
estimates. By applying perturbation-theory expressions from
Ref.~\cite{liu18prb,liu20prl} to the multi-orbital Hubbard model parametrized
by DFT calculations, we conclude that \nabaco\ features a substantial
ferromagnetic Kitaev exchange $K_1$. The competition of antiferromagnetic $J_1$
and ferromagnetic $K_1$ lowers the saturation field and the Weiss temperature,
but promotes the buildup of magnetic correlations to higher temperatures.

\begin{figure}[tb]
    \includegraphics[width=\columnwidth]{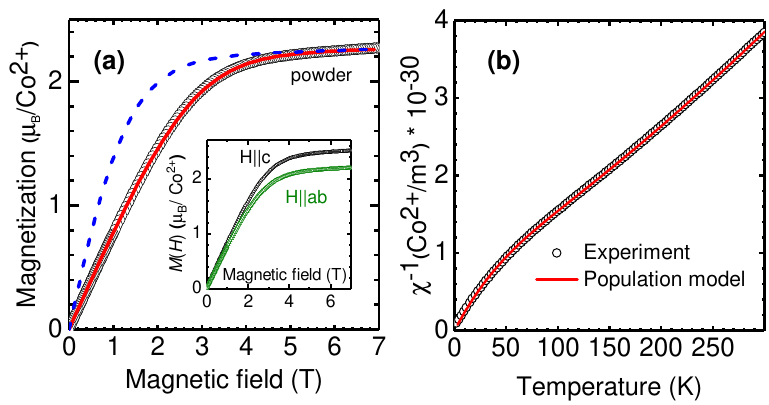}
    \caption{(a) Main panel: Powder $M(H)$ dependence at $T$\,=\,2\,K (data points). {A
	delayed increase of the $M(H)$ curve compared to the Brillouin function of
	noninteracting spins  $S = 1/2$ (dashed curve) indicates a sizable AFM
        interaction. Accounting for it, as explained in the text, yields a good
        agreement with the experiment (red solid curve)}; Inset: Single-crystalline
        $M(H)$ dependence for the ${\bf H}\parallel c$~axis and ${\bf H}\parallel
        ab$~plane. (b) Powder $\chi^{-1}(T)$ dependence at $\mu_{0}H$\,=\,20\,mT (data
        points). The solid line depicts the calculated curve according to the population
        model (see the text).}
    \label{fig:Magnetization}
\end{figure}

\section{Experimental results}
Magnetization and ESR experiments were conducted on both powder and
single-crystal samples of \nabaco\  synthesized and thoroughly characterized in
Ref.~\cite{zhong19}. Representative magnetization $M(H)$ and the inverse
susceptibility $\chi^{-1}(T) = H/M(T)$ curves are shown in
Fig.~\ref{fig:Magnetization}. 

The $M(H)$ dependence can be well fitted to the equation $M =
g\tilde{j}_{\text{eff}}B_{1/2}((H_{\text{0}} + H_{\text{ex}}),T) + \chi_{0}H_{\text{0}}$.  Here,
$B_{1/2}$ is the Brillouin function for 
spin 1/2, $g$ is the $g$~factor,
$H_{\text{0}}$ is the external magnetic field, $\chi_{0}$ accounts for the van
Vleck and diamagnetic susceptibility, and $ H_{\text{ex}} = a_{\text{ex}}M$ is
the exchange field with the parameter $a_{\text{ex}}$ characterizing the
strength of the isotropic exchange interaction $J$ =
$a_{\text{ex}}g^2\mu_{\text{B}}/N_{\text{NN}}$ with $N_{\text{NN}}$ being the number
of next neighbors. A representative fit for the powder sample is shown in the
main panel of Fig.~\ref{fig:Magnetization}(a). It yields AFM $J = 1.37$\,K and
a $g$ factor $g_\text{P}^{M}\,=\,4.38$. For the single crystal, the $g$~factors
were found to be  $g_\text{c}^{M}\,=\,4.81$ and $g_\text{ab}^{M}\,=\,4.22$. The
slight $g$~factor anisotropy is likely related to a small trigonal distortion
of the CoO$_6$ octahedra \cite{abragam70}.

\begin{figure*}[tb]
    \includegraphics[width=\textwidth]{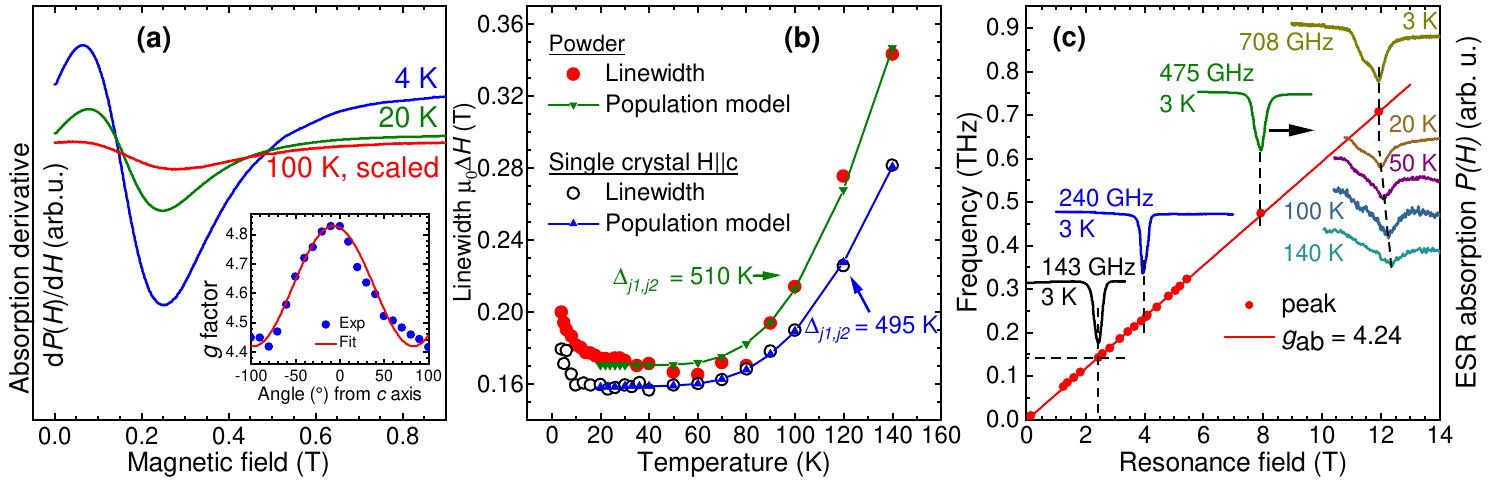}
    \caption{(a) Single-crystalline ESR signals (field derivative of the
microwave absorption) at $\nu = 9.6$\;GHz for the ${\bf H}\parallel c$~axis at
selected temperatures. Inset: Angular dependence of the $g$~factor at $\nu =
9.6$\;GHz and at $T = 4$\,K (data points). The solid line is a fit to the
dependence $g(\alpha) = [g_{\text{ab}}^2\cos^2(\alpha) +
g_{\text{c}}^2\sin^2(\alpha)]^{1/2}$. (b) $T$ dependence of the ESR linewidth
of the single- and polycrystalline samples at $\nu = 9.6$\;GHz (data points).
Solid lines are the calculated dependences according to the population model
(see the text). (c) Left scale: $\nu$ vs.\ field position of the peak of
the powder HF-ESR spectrum at $T = 3$\,K (dots). The solid line
is the fit to the dependence $h\nu = g\mu_{\text{B}}H$; Right scale: Powder
HF-ESR spectra at selected frequencies $\nu$ at $T = 3$\,K and at selected
temperatures at $\nu = 708$\,GHz.}
    \label{fig:ESR}
\end{figure*}

The inverse susceptibility $\chi^{-1}(T)$ per $\text{Co}^{2+}$ ion at
$\mu_{0}H$\,=\,20\,mT is shown in Fig.~\ref{fig:Magnetization}(b) together
with the result of a population model. The latter is derived from the
diagonalization of a single-ion Hamiltonian including the spin-orbit coupling
and the Zeeman interactions, {$\mathcal{H} = \ltilde \mathbf{S}\mathbf{L}
-\mu_{\text{B}}H (g_{\text{S}}S_{\text{z}} + g_{\text{L}}L_{\text{z}})$ with
$g_{\rm S} = 2$, $g_{\text{L}} = -3/2$} and thermal average $<M> =
\mathcal{Z}^{-1}\sum_i\bra{E_{\text{i}}}
e^{-E_{\text{i}}/k_{\text{B}}T}\mu_{\text{z}}\ket{E_{\text{i}}}$, $\mathcal{Z}$
being the partition sum. $\chi^{-1}(T)$ does not obey a simple Curie-Weiss law
$\propto T$, but rather possesses a remarkable inflection point around $T \sim
150$\,K which is reasonable to attribute to the thermally assisted population
of the excited $\tilde{j}_2 = \frac32$ multiplet (Fig.~\ref{fig:overview}).
The modeling reproduces the experimental data with the energy splitting $\Delta
E_{\tilde{j}_1,\tilde{j}_2} = E_{\tilde{j}_2} - E_{\tilde{j}_1} = 419$\,K. 

The main results of the ESR measurements for both the powder- and
single-crystalline samples at the X-band frequency of $\nu\,=\,9.6$\,GHz
conducted with a commercial ESR spectrometer from Bruker are summarized in
Figs.~\ref{fig:ESR}(a) and (b). A rather broad Lorentzian-shaped signal was
observed [Fig.~\ref{fig:ESR}(a)]. The Lorentzian fitting yielded the width
$\Delta H$ and the resonance field $H_{\text{res}}$ from which the effective
$g$~factor can be calculated according to resonance condition $h\nu =
g\mu_{\text{B}}H_{\text{res}}$.       

The angular dependent measurements revealed an anisotropy of the $g$~factor
which follows the conventional angular dependence $g(\alpha) =
\sqrt{g_{\text{ab}}^2\cos^2(\alpha) + g_{\text{c}}^2\sin^2(\alpha)}$  with
$\alpha$ being the angle which the applied field makes with the $c$~axis
[Fig.~\ref{fig:ESR}(a), inset]. The fit yielded the values
$g_{\text{c}}\,\approx\,4.83$ and $g_{\text{ab}}\,\approx\,4.42$ for the
out-of-plane and in-plane orientation of the field, respectively, consistent
with the static magnetization data. 

More accurate estimates of the $g_{\text{ab}}$ value were provided by
multi-frequency high-field ESR (HF-ESR) measurements on the powder sample which
were carried out with a homemade setup \cite{Zeisner2019}.  Selected HF-ESR
spectra at $T = 3$\,K and different excitation frequencies are shown in
Fig.~\ref{fig:ESR}(c). The signal is a single Lorentzian-shaped line at lower
frequencies and develops a shoulder at $\nu \geq 150$\,GHz. Such a structure
typical of the ESR response of a powder sample with an anisotropic $g$~factor
\cite{poole1996} can be most clearly resolved at the highest frequency of
708\,GHz. In the case of the uniaxial $g$~anisotropy the resonance field  of
the peak is determined by the $g$~factor component perpendicular to the
symmetry axis, i.e., by $g_{\text{ab}}$ in our case \cite{poole1996}. The field
position of the peak  plotted versus $\nu$ exhibits a linear
dependence according to the relation $h\nu = g_{\text{ab}}\mu_{\text{B}}H$ {and
gives $g_{\text{ab}}^{\text{HF}}\,=\,$4.24 [Fig.~\ref{fig:ESR}(c)]. Note that
this linear dependence features a negligible intercept with the frequency axis
evidencing the absence of any gap in the uniform spin excitation spectrum
larger than 9.6\,GHz (0.04\,meV). Thus, taken together the X-band and HF-ESR
results yield the {\bf g}~tensor $[g_{\text{c}},g_{\text{ab}}] = [4.83, 4.24]$
consistent with the estimate $[g^M_{\text{c}},g^M_{\text{ab}}] = [4.81, 4.22]$
from the magnetization data.  Interestingly, with increasing the temperature
close to 100\,K the peak of the HF-ESR spectrum at $\nu = 708$\,GHz shifts by
an amount of $\mu_0\delta H_{\text{0}} \approx 0.35$\,T towards higher fields,
suggesting a reduction of the $g$~factor [Fig.~\ref{fig:ESR}(c)].  At
$T\,=\,100$\,K, a linear $\nu$ versus $H$ $g$~factor fit of this peak
yields $g_{\text{c}}$(100\,K)\,=\,4.11. Such a reduction indicates the
involvement of the thermally-populated excited $\tilde{j}_2 = \frac32 $
multiplet which is characterized by a smaller $g$~factor~\cite{abragam70}. 

Both X-band and HF-ESR spectra broaden significantly with increasing the
temperature [Figs.~\ref{fig:ESR}(a) and (c)]. Analysis of the origin of
this broadening provides important insights into the spin dynamics and
temperature effects in \nabaco.  On the quantitative level, the $\Delta H(T)$
dependence can be more accurately evaluated from the X-band data because strong
fields applied in an HF-ESR experiment may cause additional inhomogeneous
broadening, and instrumental distortions of the lineshape are possible in the
HF-ESR apparatus.

In Fig.~\ref{fig:ESR}(b), the temperature dependence of the linewidth, as
obtained from the Lorentzian fits, is shown for the powder- and
single-crystalline samples.  While lowering the temperature the width decreases
down to approximately 50\,K, then remains constant until approximately 20\,K,
where it starts to rise again. The high-temperature behavior is likely due to
the above discussed population effect of the excited $\tilde{j}_2 = \frac32$
multiplet [Fig.~\ref{fig:overview}(a)]. This leads to a broadening of the
linewidth, while at lower temperatures, only the $\tilde{j}_1 = \frac12$
multiplet is populated. This behavior can be reproduced by a population model
with Boltzmann weights for the energy levels of the Co$^{2+}$ ions in an
octahedral oxygen environment, as described in Ref.~\cite{Slichter96}.  {From this
model, one obtains the energy splitting $\Delta E^{\rm
pow}_{\tilde{j}_1,\tilde{j}_2} = 510$\,K for the powder sample, which is
somewhat larger than the estimate from the modeling of the static
susceptibility. For the single crystal, one has, similarly, $\Delta E^{\rm
cr}_{\tilde{j}_1,\tilde{j}_2} = 495$\,K.} For the in-plane direction of the
field, the linewidth could not be modeled well due to the uncertainty of the
Lorentzian fits. 

The upturn of $\Delta H(T)$ below 20\,K is remarkable. It cannot be described
by the single-ion population model, suggesting that magnetic correlations
between the Co spins begin to develop gradually. Usually this signifies
proximity to a magnetic phase transition where the spin dynamics experiences a
critical slowing down \cite{dejongh90}. However, such a phase transition in
\nabaco\ takes place at two orders of magnitude lower temperature of
$T_\text{N} = 148$\,mK \cite{li20natcommun} which evidences an extended
low-temperature regime of the spin-liquid like, slow spin fluctuations in the
electron spin system.         

\section{Theory and discussion}
To understand why magnetic correlations set on at such a high temperature, we
performed a microscopic numerical analysis based on DFT band-structure as well
as multiplet calculations. For the former, we do full-relativistic nonmagnetic
DFT calculations within the generalized gradient approximation
(GGA)~\cite{PBE96}, as implemented in the full-potential local-orbital code
\textsc{fplo}~\cite{FPLO}. 

As an input, we use the experimental crystal structure described within the
space group $P\bar{3}m1$~\cite{zhong19}.  The unit cell contains one Co atom
with the local trigonal symmetry (point group $\bar{3}m$).  The GGA band
structure features a ten-band manifold crossing the Fermi level, as expected
for a unit cell containing one Co atom featuring five $d$ orbitals and two
spin flavors. In the absence of spin polarization, each band is doubly
degenerate in accord with the Kramers theorem.  As a result, the band structure
features only five distinct band dispersions (Fig.~\ref{fig:dosband}, left).
{The strong octahedral crystal field splits these bands into two manifolds that
are well separated from the rest of the valence band and are formed almost
exclusively by Co $d$ states} (Fig.~\ref{fig:dosband}, right)~\cite{Note2}.
For further analysis, we single out these states by resorting to Co-centered
Wannier functions (WFs).  Following the procedure described in
Ref.~\cite{FPLO-WF}, we obtain excellent agreement between the
Fourier-transformed WF and the respective GGA bands (Fig.~\ref{fig:dosband},
a).

\begin{figure}[tb]
    \includegraphics[width=\columnwidth]{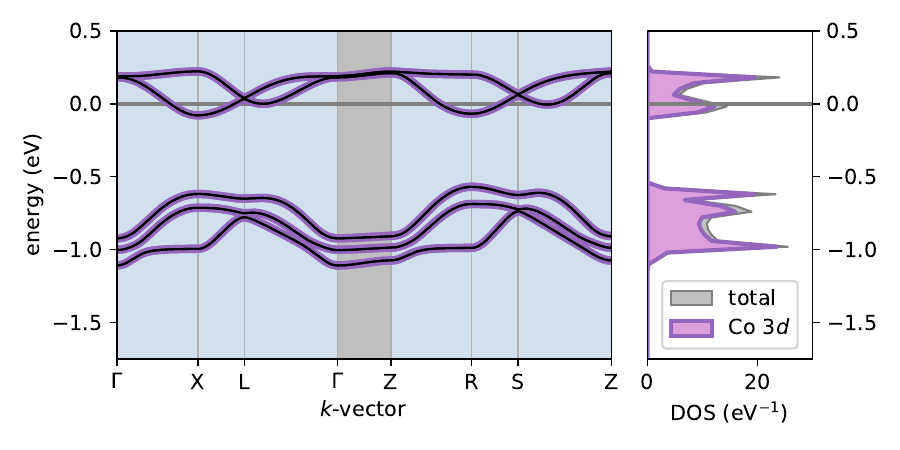}
    \caption{Left: full relativistic GGA band structure (thin black curves) and
the eigenvalues of the Fourier-transformed Wannier Hamiltonian (thick violet
curves). All bands are doubly degenerate due to the time-reversal symmetry.
Minute dispersions along $\Gamma$-Z are indicative of a quasi-two-dimensional
electronic structure. Right: total density of states (DOS) and Co $3d$
contributions. The Fermi level is at zero energy.}
\label{fig:dosband}
\end{figure}

In the WF basis, the Hamiltonian is a sum of local and nonlocal terms.  The
former describes the one-particle spectrum of an isolated trigonally distorted
CoO$_6$ octahedron, {and key parameters of the local Hamiltonian can be
extracted directly from WF}~\cite{kuzian21X}.  By inspecting different
contributions, we find a well-defined hierarchy of energy scales: the strong
cubic CF, followed by the SO interactions, the
trigonal CF splitting, and finally, the Zeeman term. The separation of energy
scales allows us to disentangle the underlying processes and arrive at an
intuitive physical picture, which we show in Fig~\ref{fig:overview}(a). First,
the cubic CF splits the ${}^4F$ multiplet of the free Co$^{2+}$ ion into three
multiplets where the lower one, ${}^4T_{1g}$, contains predominantly only
one-electron $t_{2g}$ holes. Next, the ${}^4T_{1g}$ multiplet is further split
by the SO interaction. An isolated ${}^4T_{1g}$ multiplet can be solved
analytically \cite{abragam70}, resulting in effective $\tilde j$  values, from
lowest to highest energy $\tilde j_1 = \frac12$, $\tilde j_2 = \frac32$ and
$\tilde j_3 = \frac52$ \cite{Note3}; the respective $g$ factors amount to $g_1
= \frac{13}{3}$, $g_2 = \frac{16}{15}$ and $g_3 = \frac35$, and the two
splittings amount to $\frac32\ltilde$ and $\frac52\ltilde$, {where
$\ltilde$\,=\,33\,meV is the effective SO interaction parameter}~\cite{Note4}.
The trigonal CF further splits the second and third multiplets, giving rise to
the experimentally observed $g$~factor anisotropy.  Finally, the Zeeman
interaction lifts the remaining degeneracy.

The GGA calculations (Fig.\ \ref{fig:dosband}) imply a one-electron picture in
an effective mean-field potential and cannot display the multiplet structure
directly.  To account for the ensuing electronic correlations, we perform
multiplet calculations using the parameters extracted from the WF, and obtain a
complete numerical solution of the local many-body problem. Skipping
methodological aspects of this procedure that will be published elsewhere, we
go straight to the most important result: the splitting between the $\tilde j_1
= \frac12$ ground state and the lowest-lying $\tilde j_2 = \frac32$ component
amounts to {$\Delta E_{\tilde{j}_1,\tilde{j}_2} = 472$\,K, in a fair agreement
with our  experimental estimates} \cite{SI}. The splitting exceeds by far the
magnetic energy scale.  Therefore, the magnetism of \nabaco\ can be safely
described in an effective \jhalf\ model. Another important observable is the
${\bf g}$~tensor anisotropy, i.e.\ the difference between $g_c$ and $g_{ab}$. 

Our calculations lead to an anisotropy of $g_\text{c}=4.33$ and $g_{\rm
ab}=4.55$ being slightly smaller than the experimental one and of opposite
sign. The reason for this discrepancy is the extreme sensitivity of the
$g$~factor anisotropy to the small (10\,meV) trigonal CF parameter:
a small readjustment of the CF parameter reproduces the experimental values
($g_\text{c}^{adj}=4.80$ and $g_\text{ab}^{adj}=4.31$). And since the trigonal
CF is in turn governed by the oxygen positions in the crystal structure,
additional structural refinements, especially at lower temperatures, are needed
for a conclusive analysis.

Having solved the local problem, we turn to hopping processes that underlie the
magnetic exchange. By analyzing the respective WF, we find that \nabaco\ is an
excellent model system: further longer-range in-plane as well as interplane
hoppings are orders of magnitude weaker than the leading nearest-neighbor
terms. As we are dealing with a multi-orbital problem, a direct determination
of the magnetic exchange integrals from the hopping terms is very challenging.
Therefore, we start with a qualitative analysis. By inspecting the hoppings
between different orbitals, we find that for every pair of neighboring Co
atoms, the leading contribution pertains to a single matrix element: virtual
electron transfer between two $t_{2g}$ orbitals lying in the same plane [e.g., the
plane in Fig.~\ref{fig:overview}(b)]. Although this hopping implicitly
includes transfer to and between the ligand orbitals, it is effectively
equivalent to the direct hopping denoted as $t^{\prime}$ in Refs.\
\cite{liu18prb,liu20prl}.  We note that this hopping is underlain by the
perfect mutual alignment of neighboring CoO$_6$ octahedra and may be strongly
suppressed if the octahedra are tilted. 

Next, we estimate the exchange integral $J_1$ quantitatively. To this end, we
perform full-relativistic magnetic DFT+$U$ calculations \cite{SI}.  We find
that large orbital moments, a prerequisite for the correct description of
\jhalf\  physics,  are stable only if the quantization axis coincides with the
threefold rotation crystal axis. We construct three supercells and perform
total-energy calculations for different collinear magnetic arrangements. By
mapping the total energies onto a classical Heisenberg model with effective
spins \jhalf, we estimate the nearest-neighbor exchange to be 11.6$\pm$2.0\,K,
where the error bars pertain to the uncertainty in choosing the $U$ value
($U=5\mp{}1$\,eV). 

Why is the DFT+$U$ estimate for $J_1$ so large? To find the root cause, we
first recollect that the magnetic exchange in \nabaco\ is anisotropic. As the
orientation of magnetic moments in DFT+$U$ is fixed, such calculations provide
only one component of the magnetic exchange, $J^{zz}_1$. To estimate the
anisotropic terms we go a step back and apply the recently developed
perturbative expressions, \cite{liu18prb,liu20prl} parametrized by our WF.
Using just the dominant hopping term $t^{\prime}=82$ meV~\cite{SI} we arrive at
$J^{zz}_1=J_1+K_1/3=2.3$\,K, which is smaller than the DFT+$U$ estimate.
Surprisingly, we find, besides the leading antiferromagnetic term
$J_1=(16/81)(t^{\prime})^2/U=3.1$\,K, also a sizable ferromagnetic Kitaev term
$K_1=(-4/27)(t^{\prime})^2/U=-2.3$ K.  Other anisotropic terms are smaller, and
hence nearest-neighbor exchange and Kitaev terms dominate the magnetic
interactions. {These results agree with our mean-field estimate of $J$ from the
magnetization $M(H)$ dependence} [Fig.~\ref{fig:Magnetization}(a)]. 

We are now in a position to combine our results into a coherent physical picture.
\nabaco\ is a nearest-neighbor \jhalf\  triangular magnet with a strongly
anisotropic exchange: Its magnetism is driven by the competition of the
dominant antiferromagnetic contribution with a sizable ferromagnetic Kitaev
term. The latter lowers the saturation field, which otherwise would be of the
order of $J_1$.  But the energy scale of magnetic correlations is set by the
absolute values of magnetic exchanges, and hence it is enhanced by $K_1$.
Therefore, the onset of the spin-spin correlations already at $T\sim
20$\,K\,$\gg T_{\rm N}$, manifested in the observed broadening of the ESR
linewidth, is a combined effect of $J_1$ and $K_1$.

\section{Conclusions}
Our static magnetization and ESR results on the spin-liquid candidate material
\nabaco\ enable us to classify it as an anisotropic triangular magnet in the
effective spin \jhalf\  ground state well isolated from the higher energy
effective spin-$\frac{3}{2}$ state. Furthermore, the ESR data provide strong
indications of the magnetic correlations setting in far above the magnetic
ordering temperature $T_{\rm N}$ and the Weiss temperature $\theta$. We
established the anisotropic \jhalf\ ground state on the theory level as well
and reveal, besides the antiferromagnetic isotropic nearest-neighbor exchange
interaction $J_1$,  a significant ferromagnetic Kitaev interaction term $K$,
which is beneficial for the realization of a spin liquid in this compound. The
competition between $J_1$ and $K$ reduces the temperature scale $\theta$  but
does not affect magnetic correlations the energy scale of $\sim 20$\,K\,$\gg
T_{\rm N}$ of which is probed by ESR. Altogether our findings put forward
\nabaco\ as a promising Kitaev-exchange-assisted spin-liquid material and call
for further extensive exploration of the exciting physics of this class of
compounds. 

\begin{acknowledgments}
OJ thanks Alexander Tsirlin and RH thanks Roman Kuzian for fruitful discussions
and helpful comments. We thank Manuel Richter for useful hints on DFT+$U$
calculations and Ulrike Nitzsche for technical assistance. WR and OJ were
supported by the Leibniz Association through the Leibniz Competition. This work
was supported in part by the Deutsche Forschungsgemeinschaft (DFG), grant
number KA 1694/12-1, and by the Collaborative Research Center SFB 1143
``Correlated Magnetism: From Frustration to Topology'' (Project No.
247310070). The materials synthesis was supported by the Gordon and Betty Moore
Foundation, EPiQS Program, grant number GBMF-4412.
\end{acknowledgments}

\newpage

\begin{widetext}

\begin{center}
{\large
Supplemental Material for 
\smallskip

\textbf{Frustration enhanced by Kitaev exchange in a $\tilde{\bm{j}}_{\text{eff}}=\frac12$ triangular antiferromagnet}

\medskip

\normalsize 
C.~Wellm, W.~Roscher, J.~Zeisner, A.~Alfonsov, R.~Zhong, R.~J.~Cava,\\ A.~Savoyant, R.~Hayn, J.~van~den~Brink, B.~B\"uchner, O.~Janson, and V.~Kataev}
\end{center}
\medskip

\section*{Wannier functions: local terms}
We analyse the full-relativistic non spin-polarized GGA band structure of \nabaco\ (Fig.~4 in the manuscript) by constructing Wannier functions (WF) describing the Co $3d$ states.  To this end, we resort to the trigonal basis:
\begin{equation}\label{trig_bas}
\begin{split}
	&|x\rangle=\sqrt{\tfrac{2}{3}}\,|x^2-y^2\rangle-\sqrt{\tfrac{1}{3}}\,|xz\rangle\\
	&|y\rangle=-\sqrt{\tfrac{2}{3}}\,|xy\rangle-\sqrt{\tfrac{1}{3}}\,|yz\rangle\\
	&|z\rangle=|z^2\rangle\\
	&|v\rangle=\sqrt{\tfrac{1}{3}}\,|x^2-y^2\rangle+\sqrt{\tfrac{2}{3}}\,|xz\rangle\\
	&|w\rangle=-\sqrt{\tfrac{1}{3}}\,|xy\rangle+\sqrt{\tfrac{2}{3}}\,|yz\rangle,
\end{split}
\end{equation}
In this basis, the local terms of the resulting Hamiltonian are given by the following Hermitian matrix:

\begin{equation}\label{onsite_matrix}
\small{%
\begin{array}{c}
	\begin{array}{p{.5cm}rrrrrrrrrrr}
	     & &\multicolumn{2}{c}{x} & \multicolumn{2}{c}{y}& \multicolumn{2}{c}{z}& \multicolumn{2}{c}{v}& \multicolumn{2}{c}{w}\\
	     & & \phantom{-}+\frac{1}{2}\phantom{-} & \phantom{-}-\frac{1}{2}\phantom{-} & \phantom{-}+\frac{1}{2}\phantom{-} & \phantom{-}-\frac{1}{2}\phantom{-} & \phantom{-}+\frac{1}{2}\phantom{-} & \phantom{-}-\frac{1}{2}\phantom{-} & \phantom{-}+\frac{1}{2}\phantom{-} & \phantom{-}-\frac{1}{2}\phantom{-} & \phantom{-}+\frac{1}{2}\phantom{-} & \phantom{-}-\frac{1}{2}\phantom{-}\\
	\end{array}\\
	\begin{array}{cc}
	\rule{0pt}{2.1cm}&H_0=\left(\begin{array}{*{10}{>{\centering\arraybackslash$}p{1.1cm}<{$}}}
		-793.8 & 0 & 32.9\im & 0 & 0 & -33.8\phantom{i} & \rule{4.1pt}{0pt}10.7 & \phantom{-}33.5\phantom{i} & 47.4\im & \phantom{-}33.5\im\\
		 & -793.8 & 0 & -32.9\im & \phantom{-}33.8\phantom{i} & 0 & -33.5\phantom{i} & \rule{4.1pt}{0pt}10.7 & 33.5\im & -47.4\im\\
		 &  & -793.8 & 0 & 0 & \phantom{-}33.8\im & -47.4\im & \phantom{-}33.5\im & 10.7\phantom{i} & -33.5\phantom{i}\\
		 &  &  & -793.8 & \phantom{-}33.8\im & 0 & \phantom{-}33.5\im & \phantom{-}47.4\im & 33.5\phantom{i} & \rule{4.1pt}{0pt}10.7\\
		 &  &  &  & -804.4 & 0 & 0 & -47.5\phantom{i} & 0 & \phantom{-}47.5\im\\
		 &  &  &  &  & -804.4 & \phantom{-}47.5\phantom{i} & 0 & 47.5\im & 0\\
		&&&&&&\rule{4.1pt}{0pt}88.5&0&0&0\\
		&&&&&&&\rule{4.1pt}{0pt}88.5&0&0\\
		&&&&&&&&88.5&0\\
		&&&&&&&&&\rule{4.1pt}{0pt}88.5\\
	\end{array}\right),
	\end{array}
\end{array}}
\end{equation}

where we show only the upper triangular part. All values are in meV.

\section*{Wannier functions: nearest-neighbor hopping}
To describe the nearest-neighbor hopping $t_1$, it is reasonable to resort to the standard $d$ basis ($|xy\rangle$, $|yz\rangle$, $|3z^2-r^2\rangle$, $|xz\rangle$, and $ |x^2-y^2\rangle$):

\begin{equation}\label{t1}
\begin{array}{c}
         t_1 = \left(\begin{array}{rrrrr}
-2.0	 & -17.3		 & 7.1	 & 	1.3	 & 	1.8\\
	 & 	-2.0	 & 	-2.0	 & 	1.3	 & 	7.0\\
& & 	33.8	 & 	-8.0		 & -31.2\\
& & & 	{\bf-82.8} & 	-13.8\\
& & &	 & 	-2.2\\
	\end{array}\right).
\end{array}
\end{equation}

Note that the matrix elements in Eq.~(\ref{t1}) are given for the pair of Co
atoms separated by the lattice translation vector $a$, i.e.\ along the
crystallographic direction $\left[100\right]$. We now use these hopping
parameters for a rough estimate of anisotropic exchange couplings. For that
purpose we apply the perturbative formulas which were recently developed
\cite{sm:liu18prb,sm:liu20prl} for edge shared CoO$_6$ octahedra appearing for
instance in Na$_3$Co$_2$SbO$_6$ where the Co$^{2+}$ ions build a honeycomb
lattice. These formulas can be applied in our case of Na$_2$BaCo(PO$_4$)$_2$
with the only difference that the octahedra edges do not touch directly but are
just close together such that the shortest Co-O-Co exchange path is replaced by
Co-O-O-Co. That makes, however, no difference in the perturbative formulas
which are formulated for hopping parameters between effective $d$-like
orbitals. As it was shown in \cite{sm:liu18prb,sm:liu20prl} the most relevant
hoppings are those between $t_{2g}$ orbitals. From (\ref{t1}) one can read out
a very important hopping $t^{\prime}=-82.8$ meV between neighbouring
$|xz\rangle$ orbitals and a much smaller hopping $t=-17.3$ meV between
$|xy\rangle$ and $|yz\rangle$, where we used already the notation of
\cite{sm:liu18prb,sm:liu20prl}. Taking only $t^{\prime}$ and neglecting $t$
completely we can read out from the last line of Eq. (7) in \cite{sm:liu18prb}
the isotropic exchange $J_1=(16/81) (t^{\prime})^2/U=3.1$ K and the Kitaev term
$K_1=-(4/27)(t^{\prime})^2/U=-2.3$ K.  Here, we also neglected the influence of
the trigonal splitting $\Delta$ (in the notation of \cite{sm:liu20prl}) on the
exchange terms.

\section*{Exchange integrals from DFT+$U$ calculations}
We estimate the magnetic exchanges by solving a redundant system of linear
equations parameterized by spin-polarized GGA+$U$ total energies.
Full-relativistic SGGA+$U$ calculations are done using the
fully-localized-limit double-counting correction with the onsite Coulomb
repulsion of 3.5...6.5\,eV and the onsite Hund's exchange of 1\,eV.
Figure~\ref{J_u} shows how the DFT+$U$-evaluated magnetic exchanges depend on
the Coulomb interaction $U$.

\begin{figure}
\includegraphics[width=.5\columnwidth]{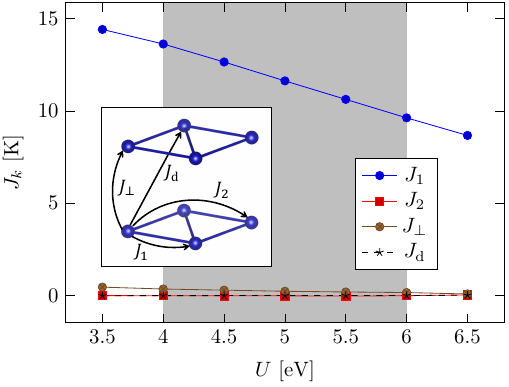}
\caption{The magnetic exchanges as a function of the onsite Coulomb repulsion $U$; the shaded area demarcates the range of plausible values. The relevant exchange pathways are schematically depicted in the inset.}\label{J_u}
\end{figure}

\section*{Multiplet calculations by exact diagonalization}

To calculate the multiplet spectrum of a single Co$^{2+}$ center we treat a
local Hamiltonian for the configuration $3d^7$, containing Coulomb, ligand
field, and spin-orbit (SO) interactions, leading to a 120*120 matrix problem.
The influence of the surrounding octahedra of O ligands with a slight trigonal
distortion (point group D$_{3d}$) is taken into account by three ligand field
(LF) parameters. The local single-ion Hamiltonian can be written as 

\begin{equation}
H=H\ts{Coul}+H\ts{LF}+H\ts{SO} \; ,
\end{equation}

and it is diagonalized exactly using the computer code ELISA (electrons
localized in single atoms) which was used before to calculate the X-ray
absorption spectra (XAS) of  Mn-TCNB (Mn-tetracyanobenzene)
\cite{sm:Giovanelli14}.  The Coulomb and spin-orbit interactions are treated as
in rotationally invariant atoms. Its matrix elements are well known, and their
parameters are fitted to the spectra for a free Co$^{2+}$ ion \cite{sm:NIST}
subject to a slight reduction in the periodic crystal.  A reduction to 80 per
cent of the free ion value gives the Coulomb Slater parameter $F^{(2)}=7.54$ eV
and $F^{(4)}=5.11$ eV, but those parameters are not very critical for the low
energy spectra. The free ion spectra are well fitted by a spin orbit coupling
of $\zeta=66$ meV (given for the one-electron basis) which coincides
accidentally with the WF parameter.

The multiplet spectrum depends crucially on the ligand field part, which can be
expressed in terms of Steven's operators. The cubic symmetry is broken due to
the elongation of the octahedron along the $c$-axis and the ligand-field (or
crystal-field) Hamiltonian is split into cubic and trigonal parts:

\begin{equation}
\label{LF}
H\ts{LF}=H\ts{cub}+H\ts{trig}
=-\frac{2}{3}B_4^0 ( O_4^0 - 20 \sqrt{2} O_4^3) +
B_2^{\prime} O_2^0 + B_4^{\prime} O_4^0 \; ,
\end{equation}

where the explicit expression of the Steven's operators is given in the book of
Abragam and Bleaney \cite{sm:abragam70}, for example.  There exist several
notations for the three ligand field parameters for $d$-electrons in the
trigonal case.  Here we use the parameters $\Delta$, $v$ and $v'$ like
Koidl~\cite{sm:koidl77} and MacFarlane~\cite{sm:macfarlane67} (see also Ref.\
\onlinecite{sm:Kuzian06}): 

\begin{equation}
B_4^0=-\frac{\Delta}{120} -\frac{1}{360} \left( v + \frac{3\sqrt{2}}{2} v^{\prime} \right)
\; , \qquad
B_4^{\prime}=-\frac{1}{140} \left( v + \frac{3\sqrt{2}}{2} v^{\prime} \right)
\; , \qquad
B_2^{\prime}=\frac{v-2\sqrt{2} v^{\prime}}{21} \; .
\end{equation}
The LF and SO parameters are determined by a fit to the on-site Wannier
expansion (\ref{onsite_matrix}). Expressing (\ref{LF}) in terms of the trigonal
basis (\ref{trig_bas}) one finds the diagonal matrix elements 
\begin{equation}
\langle x  | H_0    | x  \rangle = \langle y  | H_0    | y  \rangle = 
\frac{2}{5} \Delta + \frac{v}{3} 
\; , \qquad  
\langle z  | H_0    | z  \rangle =  
\frac{2}{5} \Delta - \frac{2}{3} v
\; , \qquad  
\langle v  | H_0    | v  \rangle = \langle w  | H_0    | w  \rangle = 
- \frac{3}{5} \Delta \; ,
\end{equation}
and the two off-diagonal elements (where we correct the misprinted sign 
of the $v^{\prime}$ term in \cite{sm:Kuzian06})
\begin{equation}
\langle x  | H_0    | v  \rangle = \langle y  | H_0    | w  \rangle = 
- v^{\prime} \; .
\end{equation}

The other off-diagonal elements of (\ref{onsite_matrix}) including the
imaginary parts correspond to the SO coupling.  So, we can simply read out the
LF and SO parameters from the matrix (\ref{onsite_matrix}) and find:

\begin{align*}
\Delta&=-885.9 \, \text{meV}  \; , & v&= 10.6 \, \text{meV} \; , \\
v^{\prime}&=-10.7 \,   \text{meV} \; , & \zeta&=67.0 \, \text{meV} \; .
\end{align*}
We remark a clear order of interactions as concerns the spin-orbit coupling and the ligand field parameter: 
$
|\Delta| \gg \zeta \gg v, v^{\prime} \; .
$
Also, the Wannier fit leads to practically the same spin-orbit coupling
$\zeta=67$ meV as that one obtained by fitting the optical spectra of a free
Co$^{2+}$ ion.  That is surprising as we expect a reduction of the SO coupling
in a crystal with respect to the free ion and can be explained by a slight
overestimation of SO coupling in the GGA functional.

Taking into account the established order of interactions, we analyze first the
multiplet spectrum neglecting the small trigonal distortion and the spin-orbit
coupling (see Table \ref{Tab2}). The level scheme follows the  general rules
for a $d^7$ configuration in octahedral environment as treated in detail in the
book of Abragam and Bleaney \cite{sm:abragam70}.  The lowest $^4 F$ multiplet
of the free ion is split into $^4 T_{1g} $, $^4 T_{2g} $ and $^4 A_{2g} $,
where $^4 T_{1g} $ is the lowest one for $\Delta < 0$ (octahedral environment).
As can be seen in Tab.\ \ref{Tab2}, the  $^4 A_{2g} $ multiplet is higher than
the lowest state with spin 1/2 ($^2 E_g $) in the given case.

\begin{table}
\begin{center}
\begin{tabular}{|c|c|c|} 
\hline
\hline
term  & degeneracy    & energy (eV)  \\  
\hline

$^4 T_{1g}$  &12 & 0.00  \\
$^4 T_{2g} $  &12 & 0.77  \\
$^2 E_g $  & 4 & 1.03  \\
$^4 A_{2g} $  & 4 & 1.66  \\
$^2 T_{1g} $  &6 & 1.72  \\
$^2 T_{2g} $  & 6 & 1.76  \\
$^4 T_{1g} $  &12 & 2.09  \\

\hline
\hline
\end{tabular}
\caption{Predicted multiplet energies of Co$^{2+}$ in Na$_2$BaCo(PO$_4$)$_2$.
The Coulomb parameters $F^{(2)}=7.54$ eV and  $F^{(4)}=5.11$ eV are chosen to
have 80 per cent of their free ion value and the cubic crystal field splitting
$\Delta=-0.8856$ eV of the Wannier fit is used. For the sake of clarity we
neglect the spin-orbit coupling $\zeta$ and the small trigonal LF parameters
$v$ and $v^{\prime}$. }
\label{Tab2}
\end{center}
\end{table}

Introducing the spin-orbit coupling of $\zeta=66$ meV leads to a splitting of
the ${}^4 T_{1g} $ multiplet.   In pure cubic symmetry the ${}^4 T_{1g}$
multiplet can be described by an effective orbital moment $L\ts{eff}=1$.
Correspondingly, the spin-orbit splitting leads to an effective $\tilde
j_1=1/2$ lowest doublet, followed by a $\tilde j_2=3/2$ quartet and a
high-lying $\tilde j_3=5/2$ multiplet. In the ideal cubic case and neglecting
any interaction to higher lying multiplets, the lowest doublet (effective spin
1/2) has an isotropic $g$-factor of $g=4.33$.  The trigonal distortion, i.e.,
the parameters $v$ and $v^{\prime}$, lead to anisotropic $g$-factors as
presented in Table \ref{Tab3}. The WF parameter values $v=10$ meV and
$v^{\prime}=-10$ meV lead to an opposite anisotropy  $\Delta
g=g_c-g_{ab}=-0.22$  with respect to the experimental results.  But a slight
correction recovers the correct order. That correction is not unique since $v$
and $v^{\prime}$ influence the $g$-factor anisotropy in the form $\Delta g=(50
v + 73 v^{\prime})/ \mbox{eV}$ when all the other parameters are fixed and we
present in Table \ref{Tab3} two parameter sets which lead to $\Delta g$ being
in reasonable agreement with experiment.  We give in Table \ref{Tab3} also the
trigonal splitting parameter $\Delta_{\rm{Lines}}$ which was introduced by
Lines \cite{sm:lines63} and used recently \cite{sm:liu20prl} to analyze
Na$_3$Co$_2$SbO$_6$. It is defined as the trigonal splitting in the ${}^4
T_{1g}$ multiplet when spin-orbit coupling is switched off. 

\begin{table}
\begin{center}
\begin{tabular}{| c | c| c   |  c |  c | c   |} 
\hline
\hline
$v$ (meV)  & $v^{\prime}$ (meV)   & $g_c $ & $g_{a/b}$ & $\Delta g$ & $\Delta_{Lines}$ (meV)\\  
\hline
0 & 0 & 4.48 & 4.48 & 0.0 & 0.0\\
10 & -10 & 4.33 & 4.55 & -0.22 & -2.1\\
10 & 0 & 4.81 & 4.31 & 0.50 & 9.6\\
4 & 4 & 4.80 & 4.31 & 0.49 & 8.5\\
\hline
\hline
\end{tabular}
\caption{$g$-factors for $\zeta=66$ meV and varying trigonal distortion.}
\label{Tab3}
\end{center}
\end{table}

The SO coupling parameter $\zeta$ for one $d$ electron is related to
$\lambda=-\zeta/3$ being the SO coupling between $S=3/2$ and $L=3$ in the ${}^4
F$ multiplet of a free ion. The effective SO coupling constant $\tilde
\lambda=- 3 \lambda/2$ is defined as the coupling between $S=3/2$ and
$L_{\rm{eff}}=1$ in cubic symmetry.  The theoretical SO coupling $\zeta=66$ meV
leads to the low-energy scheme which is shown in Table~\ref{Tab4} with an
energy difference $\Delta E_{\tilde j_1 \tilde j_2}=40.7$ meV (or 472 K)
between the lowest $\tilde j=1/2$ doublet and first excited quartet $\tilde
j=3/2$. As already noted in the main text, this theoretical energy difference
lies in between the energy splittings deduced from the susceptibility data (419
K) and that one from the broadening of the ESR line width (510 K for the powder
sample). The susceptibility data correspond to a SO coupling of $\zeta=48.1$
meV (deduced from $\Delta E_{\tilde j_1 \tilde j_2}= 3 \tilde \lambda /2 = 3
\zeta / 4$) and we show in Table \ref{Tab4} also the low-energy spectrum for
this SO coupling parameter.

\begin{table}
\begin{center}
\begin{tabular}{| c | c| c   |   }   
\hline
\hline

    & $\zeta=66.0$ meV &  $\zeta=48.1$ meV \\   
\hline
$\tilde j=1/2$ & 0.0 meV & 0.0 meV \\
\hline
$\tilde j=3/2$ & 40.7 meV & 30.0 meV \\
              & 45.6 meV & 34.7 meV \\
\hline
$\tilde j=5/2$ & 112.4 meV & 84.0 meV \\              
              & 113.6 meV & 85.7 meV \\
              & 128.3 meV & 93.7 meV \\
\hline
$g_c$       &    4.80  &  4.86 \\
$g_{a/b}$ &    4.31   &  4.21 \\

\hline
\hline
\end{tabular}
\caption{Comparison of energy splitting in the lowest multiplet  ${}^4 T_{1g} $
and of the $g$-factors for two different spin-orbit coupling parameters
$\zeta$. The trigonal LF parameters were chosen to be $v=v^{\prime}=4$ meV. }
\label{Tab4}
\end{center}
\end{table}

\end{widetext}
\end{document}